\documentclass[structabstract]{aa}  
\usepackage{graphicx}
\usepackage{natbib}
\usepackage[varg]{txfonts}
\usepackage{color}
\usepackage{longtable,lscape}
\usepackage{textcomp}
\usepackage{deluxetable}
\usepackage{epsfig}

\newcommand{\los}{\emph{los}}
\begin{document}
\title{Exploring the total Galactic extinction with SDSS BHB stars}
\author{Hai-Jun Tian\inst{1,2,3}, Chao Liu\inst{1},  Jing-Yao Hu\inst{4}, Yang Xu\inst{3,4}, and Xue-Lei Chen\inst{1,5}}
\institute{
1. Key Laboratory of Optical Astronomy, National Astronomical Observatories, Chinese Academy of Sciences, Beijing 100012\\
2. LAMOST Fellow\\
3. China Three Gorges University, Yichang, 443002\\
4. National Astronomical Observatories, Chinese Academy of Sciences, Beijing, 100012\\
5. Center of High Energy Physics, Peking University, Beijing, 100871
}

 
\abstract
    {}
    {We used 12,530 photometrically-selected blue horizontal branch (BHB) stars from the Sloan Digital Sky Survey (SDSS) to estimate the total extinction of the Milky Way at the high Galactic latitudes, $R_V$ and $A_V$ in each line of sight. }
   {A Bayesian method was developed to estimate the reddening values in the given lines of sight. Based on the most likely values of reddening in multiple colors, we were able to derive the values of $R_V$ and $A_V$.}
   {
We selected 94 zero-reddened BHB stars from seven globular clusters as the template. The reddening in the four SDSS colors for the northern Galactic cap were estimated by comparing the field BHB stars with the template stars. The accuracy of this estimation is around 0.01\,mag for most lines of sight. We also obtained $<R_V>$ to be around 2.40$\pm1.05$ and $A_V$ map within an uncertainty of 0.1\,mag. The results, including reddening values in the four SDSS colors, $A_V$, and $R_V$ in each line of sight, are released on line. In this work, we employ an up-to-date parallel technique on GPU card to overcome time-consuming computations. We plan to release online the C++ CUDA code used for this analysis.  
 }
   {The extinction map derived from BHB stars is highly consistent with that from Schlegel, Finkbeiner \& Davis(1998). The derived $R_V$ is around 2.40$\pm1.05$. The contamination probably makes the $R_V$ be larger.}
   \keywords{ISM: dust, extinction -- Stars: horizontal-branch -- Galaxy: stellar content -- methods: statistical}

\authorrunning{Hai-jun Tian, Chao Liu, etc.}

  \maketitle

\section{Introduction}\label{intro}

Our Galaxy is full of interstellar medium (ISM), such as gas and dust grains \citep{draine03}, which are produced by the nuclear burning within the stars and blown out by supernova explosions and stellar winds.  The absorption and scattering of the starlight emitted from distant objects by the dust grains leads to the Galactic interstellar extinction, which varies with directions depending of the different composition of the dusts in the interstellar space. The interstellar extinction is also referred to as the reddening, or the color excess, because the tendency of the absorption and scattering is much larger in the blue than in the red wavelength. The extinction as a function of the wavelength is related to the size distribution and abundances of the grains. 
Therefore, it plays an important role in understanding the nature of the interstellar medium.

The flux of extragalactic objects, such as galaxies, quasars, etc., suffers different extinction in different bands. This effect leads to some bias on the extragalactic studies \citep{guy10,ross11, Fang11, tian11}. Therefore, understanding the total interstellar extinction in every line of sight is crucial for accurate flux measurements. The all-sky dust map can either be constrained by measuring interstellar extinction, or constrained by employing a tracer as ISM, e.g., HI. One of the most broadly used dust maps was published by \cite{SFD98} (hereafter SFD), which was derived from observations of dust emission at 100\,$\mu$m and 240\,$\mu$m with an angular resolution at about 6\,arcmin.  Since then, many other works have claimed discrepancy with their results. \cite{Stanek98} inferred that SFD overestimated the extinction in some large extinction regions from the study of the globular clusters. \cite{arce99} argued that the SFD dust map overestimated  by 20-40\% in Taurus region from star counts, colors, etc. \cite{Dobashi05} studied the optical star counts and concluded that SFD overestimated the extinction value by at least a factor of two. \cite{peek10} independently estimated the extinction using the standard galaxies and their result is consistent with the SFD map in most of the area within the uncertainty of 3\,mmag in $E(B - V)$, though some regions do deviate from SFD by up to 50\%. \cite{schlafly10}, \cite{schlafly11}, and \cite{yuan13} use the blue tip of Sloan Digital Sky Survey (SDSS) stellar locus, SDSS stellar spectra , and multiple passbands from several photometric imaging surveys respectively to measure the reddening and claim SFD overestimates $E(B-V)$ by about $14\%$. However, \cite{berry11} argues that the overestimation only exists in the southern sky, while the dust map at high northern Galactic latitudes looks good.

In this work, we use the blue horizontal branch stars (BHB) to map the total interstellar extinction over the northern Galactic cap covered by the SDSS survey. The BHB stars are luminous and far behind the dust disk, which contribute to most of the interstellar extinction. Thanks to the five-band high accuracy photometry provided by SDSS, the BHB stars are available to derive the reddening in multiple color indexes which allows us to estimate the extinction law, or $R_V$, simultaneously.

The outline of this paper is as follows. In Sect. 2, we describe the Bayesian method for the estimation of the total extinction via BHB stars and the estimation of the mean $R_V$ and $A_V$ in the given lines of sight (hereafter \los). In section 3, we assess the performance of the Bayesian method for the total extinction, and validate the method to estimate the $R_V$. In section 4, we describe the criteria to select the BHB sample. We selected a total of 12,530 BHB stars with these criteria, covering most of SDSS footprint. The result of the color excess in four color indexes and the comparisons with other works are presented and discussed in section 5. In the last section, we draw the conclusions.
\section{Method}
\label{method}

\subsection{Bayesian method for color excess}

The total Galactic extinction in a given \los\ is measured from the offset of the measured color indexes of the BHB stars from their intrinsic color.  A set of BHB stars,  whose dereddened color index, $\{\bf{c_{k}}\}$ (where $k=1, 2,\dots,N_{BHB}$), is around their intrinsic color index, are selected as template stars. The reddening of a field BHB stars can then be estimated by comparing measured colors with the templates. Given a \los\ $i$ with $N_i$ field BHB stars, the posterior probability of the reddening $\bf{E_i}$ is denoted as $p(\bf{E}_{i}|\{\bf{\hat{c}_{ij}}\},\{\bf{c_k}\})$, where ${\bf \hat{c}_{ij}}$ is the observed color index vector of the BHB star $j$ in the \los\ $i$, and $\bf{c_k}$ is the intrinsic color index vector of the template BHB star $k$. According to the Bayes theorem, this probability can be written as
 
\begin{equation}\label{eq:bayes}
p({\bf E_{i}}|\{\bf{\hat{c}_{ij}}\},\{\bf{c_k}\}) = p(\{\bf{\hat{c}_{ij}}\}|\bf{E}_{i}, \{\bf{c_k}\})P({\bf E}_{i}|\{\bf{c_k}\}).
\end{equation}

Since ${\bf E}_{i}$ is independent of the template BHB stars' $\{\bf{c_k}\}$, the prior $P({\bf E}_{i}|\{\bf{c_k}\})$ is equivalent to $P({\bf E}_{i})$, which is assumed to be flat in this work.
The right-hand side of Eq.~\ref{eq:bayes} can now be rewritten as
\begin{equation}
p({\bf E_{i}}|\{\bf{\hat{c}_{ij}}\},\{\bf{c_k}\})  =  \prod_{j=1}^{N_{i}} \sum_{k=1}^{N_{BHB}} p({\bf \hat{c}}_{ij}|{\bf E}_{i}, {\bf c}_{k})p({\bf E}_{i}).
\end{equation}

We assume that the likelihood $p({\bf \hat{c}_{ij}}|({\bf E_{i}}, \bf c_{k}))$ is a multivariate Gaussian  and so can be expressed as :

\begin{equation}
p({\bf \hat{c}}_{ij}|{\bf E}_{i}, \bf c_{k}) =  \frac{1}{(2\pi|{\bf \Sigma}|)^{m/2}}\exp(-{\bf x}^{T}{\bf \Sigma}^{-1}{\bf x}),
\end{equation}
where ${\bf x} = {\bf E} + {\bf c}_{k} - \hat{{\bf c}}_{ij}$,  and ${\bf \Sigma}$ is the m$^{th}$ rank covariance matrix of the measurement of the color indexes of the star $j$; ${\bf \Sigma}$ is composed of the measurement error
\begin{equation}
{\bf \Sigma}  =   \left[ \begin{array}{cccc}
\sigma_{u}^{2} + \sigma_{g}^{2} & -\sigma_{g}^{2}& 0 & 0 \\
-\sigma_{g}^{2} & \sigma_{g}^{2} + \sigma_{r}^{2} & -\sigma_{r}^{2} & 0 \\
0 & -\sigma_{r}^{2} & \sigma_{r}^{2} + \sigma_{i}^{2} & -\sigma_{i}^{2}\\
0 & 0 & -\sigma_{i}^{2} & \sigma_{i}^{2} + \sigma_{z}^{2}
\end{array} \right],
\end{equation} 
where the $\sigma_{u}$,$\sigma_{g}$, $\sigma_{r}$, $\sigma_{i}$, and $\sigma_{z}$ are the uncertainties of magnitudes in the $u$, $g$, $r$, $i$, and $z$ bands, respectively.

\subsection{$R_V$  and $A_V$}\label{RvAvmethod}
After deriving the probability of the reddening in a \los, the most likely reddening values, 
\begin{equation}
E_i=(E(u-g),E(g-r),E(r-i),E(i-z)),
\end{equation} 
can be obtained from the probability density function (PDF). They can be used to derive the $R_V$ and $A_V$ given an extinction model, such as the one in \cite{CCM89} (hereafter CCM),
\begin{eqnarray} \label{Extinc}
E(u-g) = ((a_{u} + \frac{b_{u}}{R_{V}}) - (a_{g} + \frac{b_{g}}{R_{V}}))*A_{V} \nonumber\\
E(g-r) = ((a_{g} + \frac{b_{g}}{R_{V}}) - (a_{r} + \frac{b_{r}}{R_{V}}))*A_{V} \nonumber\\
E(r-i) = ((a_{r} + \frac{b_{r}}{R_{V}}) - (a_{i} + \frac{b_{i}}{R_{V}}))*A_{V} \nonumber\\
E(i-z) = ((a_{i} + \frac{b_{i}}{R_{V}}) - (a_{z} + \frac{b_{z}}{R_{V}}))*A_{V}.
\end{eqnarray}
These are linear equations for $A_V$ and $A_V/R_V$ and can be easily solved with a least-squares or $\chi^2$ method to find the values of $A_V$ and $R_V$ for each BHB star. The $A_V$ and $R_V$ values in each \los\ are obtained by the median value of all the BHB stars located in each \los; the errors can be estimated by the median absolute deviation. For the wavelength of the five bands ($u$, $g$, $r$, $i$, and $z$) in SDSS, we adopt 3551$\rm\AA$, 4686$\rm\AA$, 6165$\rm\AA$, 7481$\rm\AA$, and 8931$\rm\AA$ as their effective wavelengths, respectively \citep{Fukugita96, Stoughton02}, $a_x$ and $b_x$ are calculated based on CCM.




\subsection{Template BHB stars}
To estimate the total extinction for a given \los, we need to find a set of BHB samples as the template, i.e., those having accurate color indexes without extinction. Metal-poor globular clusters are good sample for extracting this kind of template BHB stars. There are seven globular clusters covered by SDSS in the northern Galactic cap. Their basic parameters, distance ($R_{sun}$), metallicity ([Fe/H]), color excess (E(B-V)), etc., are selected from the literature \citep{harris96}. We cross-identify the stars located within radius of 30 arcmin around the globular clusters with the BHB catalogue \citep{harris96} . Then we remove the field stars whose magnitudes in the $g$ band are different by more than 0.5\,mag from the corresponding g-band magnitude of the horizontal branch of a globular clusters. Finally, 94 BHB stars are remained. We validated the 94 template stars according to their distances calculated using Eq. 5 in \cite{Fermani13} with the known metallicity and the $g-r$ of each globular cluster \citep{harris96}. A BHB star whose distance is different by more than 5$kpc$ from the corresponding globular cluster is probably not located in this globular cluster. According to this criterion, all the 94 BHB stars are within their corresponding globular cluster. 

Based on the CCM algorithm, Doug Welch\footnote{http://dogwood.physics.mcmaster.ca/Acurve.html} provides an absorption law calculator to determine the total absorption at wavelengths between 0.10 $\mu m$ and 3.33 $\mu m$. Given $R_{V}=3.1$, $A_V$, and the average wavelength of the $u, g, r, i$, and $z$ bands, we obtain the total extinction in each band:
\begin{eqnarray}\label{ebv2A}
A_u &=& 4.896*E(B-V) \nonumber\\
A_g &=& 3.788*E(B-V) \nonumber\\ 
A_r &=& 2.728*E(B-V) \nonumber\\
A_i &=& 2.093*E(B-V) \nonumber\\
A_z &=& 1.503*E(B-V).
\end{eqnarray}
The reddening  in each color can now be obtained by
\begin{eqnarray}\label{A2Eab}
E(u-g) &=& A_u - A_g \nonumber\\
E(g-r) &=& A_g - A_r \nonumber\\ 
E(r-i) &=& A_r - A_i \nonumber\\
E(i-z)&=& A_i - A_z.
\end{eqnarray}

Table \ref{tbl-1} provides the basic parameters, including $E(B-V)$, for all the seven globular clusters. Equation \ref{ebv2A} is applied to convert the $E(B-V)$ to total extinction in each band. The $E(a-b)$ listed in the table are estimated for the case of $R_V=3.1$. Alternative $R_V$ values do not significantly change the reddening of the template stars. For instance, the variation of $E(a-b)$ ($a$ and $b$ are two arbitrary SDSS bands) is less than 5\textperthousand\ when $R_V$ changes from 2.5 to 4.1. This is because the $E(B-V)$ for the clusters are very small and the difference in the reddening due to the variate extinction law is negligible owing to the accuracy of the photometry in SDSS. The intrinsic colors of the template BHB stars is then obtained by subtracting the small reddening in each color.

\begin{figure*}
\epsscale{1.0}
\plotone{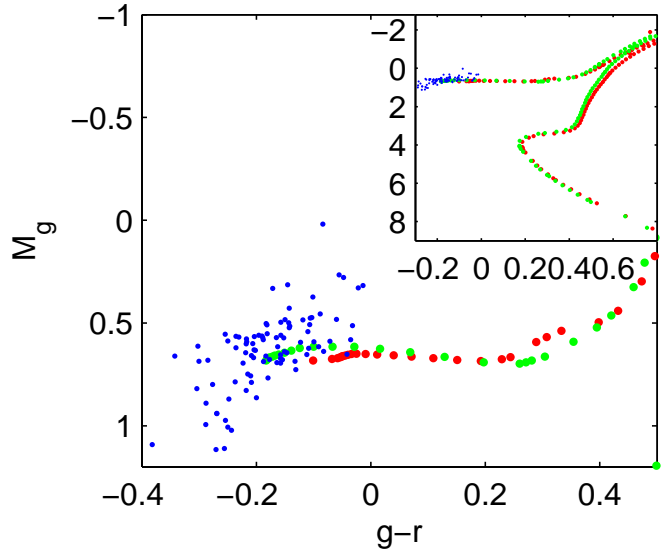}
\caption{The 94 dereddened BHB stars and two typical isochrones in the  color vs. absolute magnitude plane. The blue points are the BHB stars, the green curve is the isochrone with metallicity of -2.28\,dex and age of 13.5\,Gyrs. And the red curve is another isochrone with metallicity -2.0 at the same age. Both isochrones are obtained from \citet{girardi10}. 
\label{Color_Mg}}
\end{figure*}

Figure ~\ref{Color_Mg} shows the 94 dereddened BHB stars and two typical isochrones in a color-absolute magnitude diagram. The blue points are the BHB stars. The green and red curves are the isochrones with age 13.5\,Gyr and metallicity -2.28 and -2.0\,dex, respectively\footnote{http://stev.oapd.inaf.it/cgi-bin/cmd\_2.3} \citep{girardi10}. It shows that the observed BHB templates are consistent with the isochrones.

\section{Validation of the method}

At high Galactic latitudes, the total extinction is usually very small, even smaller than 0.1\,mag in E(B-V). Figure ~\ref{Color_Mg} shows that the template BHB stars have dispersion up to 0.3\,mag in $g-r$ because of the spread in effective temperatures, and so it is important to examine the validity of our method and to assess the performance of the reddening estimation before applying the method to the real data. We ran Monte Carlo simulations based on the template BHB stars. We supposed that each simulation was for one \los, i.e., all tracers have the same reddening values but different photometric uncertainties in each \los. For each \los, we selected a random reddening in the $u$, $g$, $r$, $i$, and $z$ bands for arbitrarily selected BHB template stars with random Gaussian noises to mimic the uncertainties of the photometry. 
Specifically, an observed color index in the simulation is obtained by
\begin{equation} \label{gaussion}
c=c_0+\epsilon+E(c) ,
\end{equation}
where $c_0$ is the noise-free, dereddened color index, $e$ is the random error following a Gaussian with zero mean and $\sigma\in[0.02,0.08]$, and $E(c)$ is the reddening.

In this work we concentrate on the extinction at high Galactic latitudes, and so we only test the low extinction cases. Considering that $R_V$ may be variant with the \los, we let the extinction curve have a freedom of changing, specifically, the reddening value $E(u-g)=E(g-r)+\Delta_{ug}$; both the $E(g-r)$ and $\Delta_{ug}$ are arbitary values for each \los, and $E(g-r) \in [0, 0.14]$\,mag, $\Delta_{ug}\in [0, 0.05]$. We arbitrarily selected 10--50 BHB template stars as field stars in each \los. 

\begin{figure*}
\epsscale{1.5}
\plotone{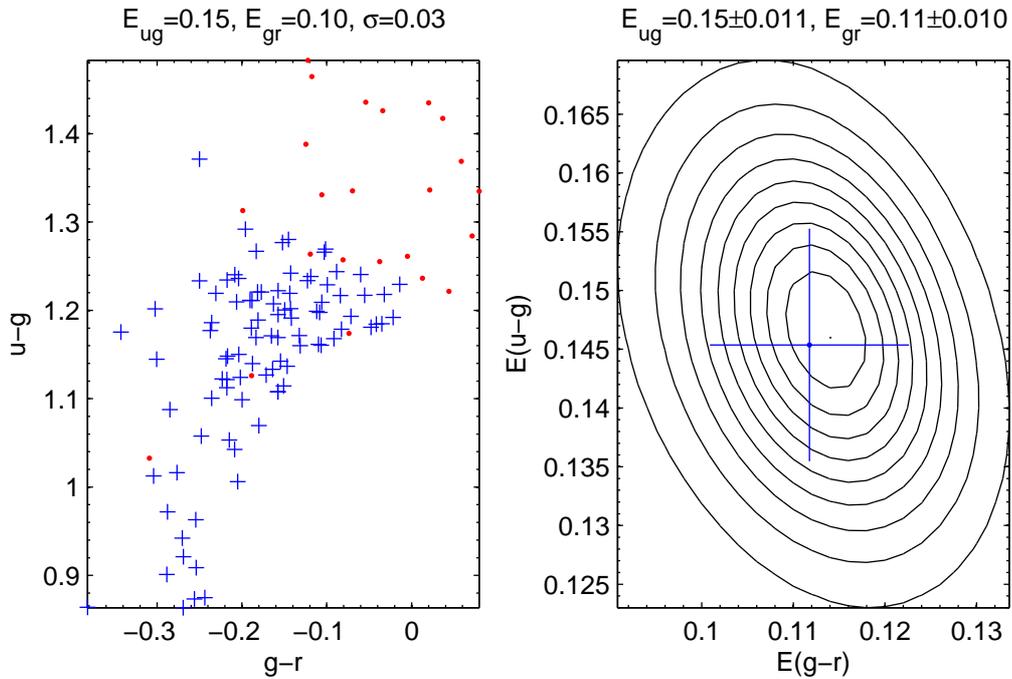}
\caption{One simulated \los\ example of 900 Monte Carlo simulations for validation of the Bayesian method. In the left panel, the blue crosses stand for the 94 template BHB stars; the red points are selected field stars with random reddening values and random Gaussian noises in photometry. The right panel shows the contour of the posterior probability of the reddening $E(u-g)$ and $E(g-r)$ estimated from the Bayesian method. The large blue cross marks the most likely reddening value with a $1\sigma$ error.
\label{fig_simu1}}
\end{figure*}
\begin{figure*}
\epsscale{1.5}
\plotone{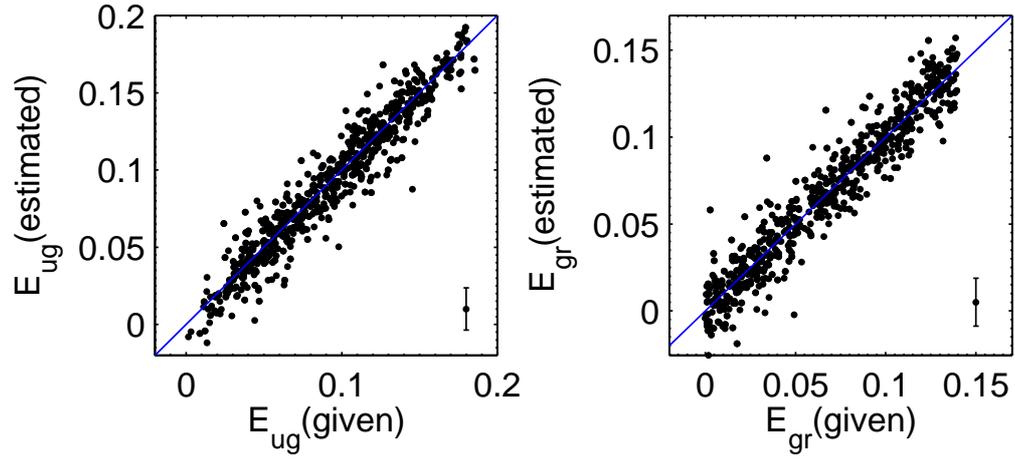}
\caption{The estimated reddening values vs. the given values in the simulation. The mean 1$\sigma$ error bars are marked at the bottom.
\label{fig_simu_comparing}}
\end{figure*}
\begin{figure*}
\epsscale{1.0}
\plotone{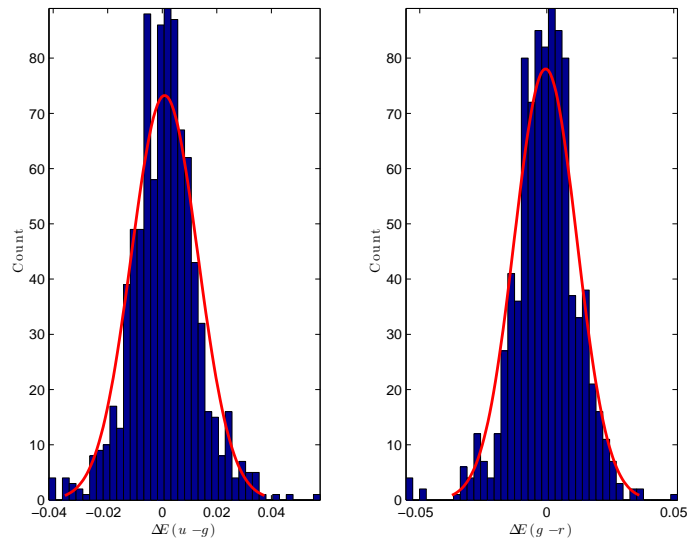}
\caption{ Histogram of the residuals of the estimated reddening values in the simulation. The red curves are the Gaussian fit profiles with $\sigma \simeq 0.01$.
\label{fig_simu_contrast}}
\end{figure*}

We applied the method described in section 2 to the mock stars and ran the Monte Carlo simulations for 900 \los\ with different testing samples. One of the simulations is presented in Fig.~\ref{fig_simu1}. And Fig.~\ref{fig_simu_comparing} shows the comparison between the estimated and the true extinction values in four colors. Figure ~\ref{fig_simu_contrast} shows the histogram of the residuals of the estimated values. It is noted that most of the residuals are less than 0.01\,mag, which suggests that the Bayesian method we employed in this work is robust.

\begin{figure*}
\epsscale{0.6}
\plotone{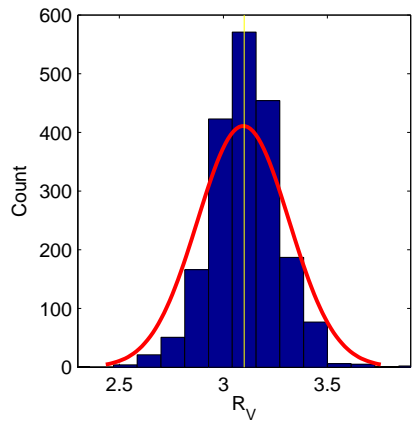}
\caption{Histogram of estimated $R_V$ in the simulation. The red curves are the Gaussian fit profiles with the mean value $<R_V> \simeq 3.1$ and $\sigma \simeq 0.16$; the yellow line marks the location of $R_V = 3.1$. \label{Rv_Hist_sim}}
\end{figure*}

To validate the method of the $R_V$ estimation, we initially obtained the $E(B-V)$ values in each \los\ from the extinction maps released by SFD.  The values of $E(u-g)$ and $E(g-r)$ is estimated according to Eqs.~\ref{ebv2A} and~\ref{A2Eab}, given $R_V=3.1$. We then solve Eq.~\ref{Extinc} with the least-squares or $\chi^2$ method described in section~\ref{RvAvmethod} to obtain the $R_V$ in each \los. Figure \ref{Rv_Hist_sim} displays the histogram of $R_V$ distribution; the yellow vertical line marks the location of $R_V=3.1$. The peak of $R_V$ distribution is exactly on the yellow line, proving the accurate performance of the method for the $R_V$ estimation.

\section{Data}

Sloan Digital Sky Survey has provided an uniform and contiguous imaging of about one third of the sky, mostly at high Galactic latitudes \citep{aihara2011}. The SDSS imaging is performed nearly simultaneously in the five optical filters: $u$, $g$, $r$, $i$, and $z$ \citep{gunn98, Fukugita96}. The images are uniformly reduced by the photometric pipeline, and the completeness of data reduction is around 95\% under the limited magnitudes 22.1, 22.4, 22.1, 21.2, and 20.3 for the five bands, respectively.

The BHB sample used in this work was selected from \cite{smith2010}, the authors identified 27,074 BHB stars candidates out of 294,652 stars from SDSS Data Release 7 (DR7) using the support vector machine (SVM) trained by the spectroscopic sample of BHB stars \citep{xue08}. 

\begin{figure*}
\epsscale{1.2}
\plotone{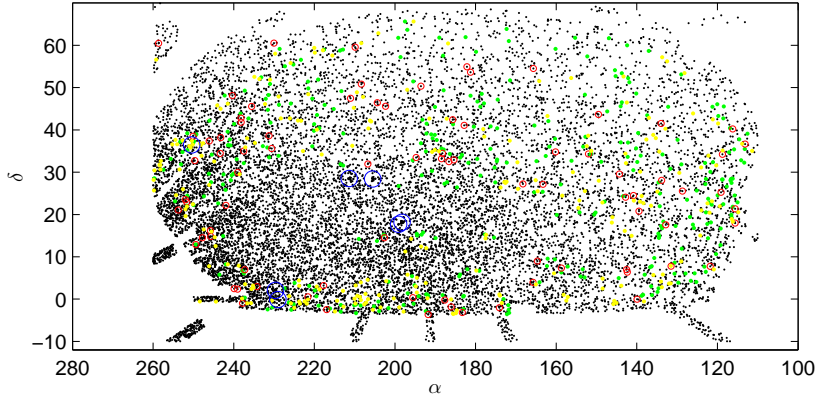}
\caption{Data sample distribution. The black points are the 13,143 BHB candidates selected in this work. We confirm that these samples are contaminated by at least 76 main-sequence stars (red circles), 311 blue straggler (green points), and 227 other objects (yellow points)\citep{xue08}. The seven globular star clusters listed in Table \ref{tbl-1} are marked by blue circles.\label{fig_data}}
\end{figure*}

Most of the BHB stars in the northern Galactic cap region are more distant than 10 kpc, as shown in the Fig. 19 of  \cite{smith2010}. Since the integrated line-of-sight extinction mainly comes from the Galactic gas disk, the typical thickness of the gas disk is far less than 10 kpc; therefore, the distant BHB stars are suitable for the estimation of the total Galactic extinction. In order to trace the extinction in most of the SDSS \los, we selected a total of 12,530 field BHB stars according to the following criteria:
\begin{itemize}
    \item[--] $g <19.0$, the BHB stars are bright enough, which ensures that  i) there is less contamination and ii) the photometric error is sufficiently small (on average smaller than 0.020 mag in the SDSS bands).
    \item[--] the sky coverage to $110^\circ<\alpha<260^\circ$ and $-10^\circ<\delta<70^\circ$, where the sky is continuously covered by SDSS DR7.
   \item[--] $(g-r)>-0.3$, according to \cite{yanny2000}.
   \item[--] $P_{SVM}>0.5$,  $P_{SVM}$ is the BHB probability from the SVM provided by \cite{smith2010}. The average contamination and completeness are about 30-40\% and 70-80\% when the classification threshold is equal to 0.5, according to Figs. 11 and 12 in \cite{smith2010}.
   \item[--] remove the contamination, such as blue stragglers (BS), main-sequence (MS) stars, etc. (as shown in Fig. \ref{fig_data}), which have been ambiguously classified by \citet{xue08}. Even then, most of the contaminations cannot be removed because we do not have the spectra for the entire BHB sample.
\end{itemize}

\begin{figure*}
\epsscale{1.6}
\plotone{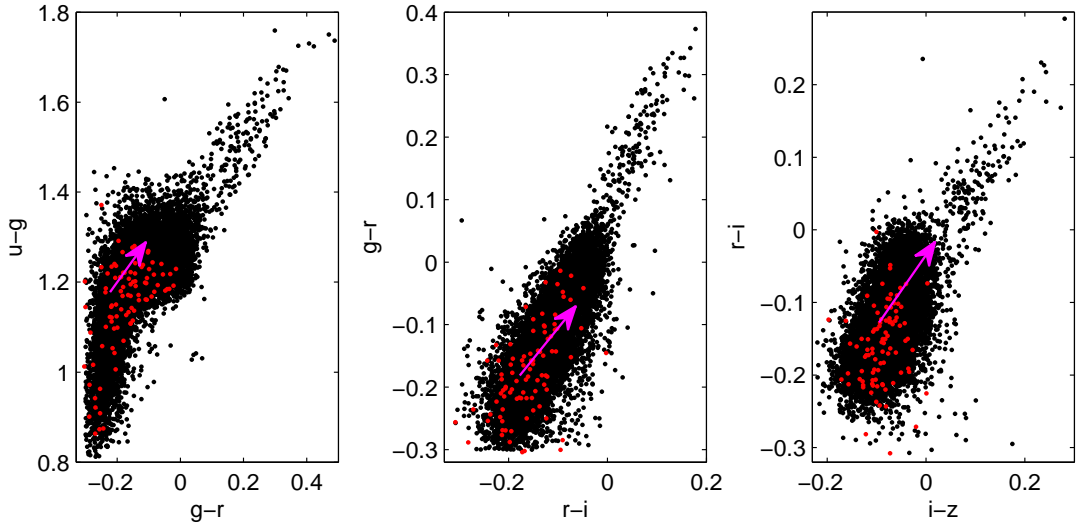}
\caption{The panels show all of the BHB stars used in this work in the color-color space. The tails of the black points are due to the extinction. The red points are the 94 template BHB stars calibrated with the known extinction. The magenta arrows show the reddening direction, the starting point of each arrow is the median value of the color indexes of the 94 template stars, and the terminal is the reddened original point with $R_V = 3.1$ and $E(B-V)=0.1$.
\label{fig_data_color}}
\end{figure*}

Figures~\ref{fig_data} and~\ref{fig_data_color} present the selected data distributions in the equatorial coordinate and color-color spaces, respectively.

\section{Analysis and results} \label{bozomath}
The sky is separated into $1^\circ\times1^\circ$ \los. The nearest 18 BHB stars around the center of the \los\ are involved in the estimation of the reddening. However, at high Galactic latitudes, the BHB stars are too sparse and extend to larger areas. In these \los, the spatial resolution of the reddening is lower than the average level. Unlike the extinction near the Galactic disk, the extinction is usually very low at high Galactic latitudes and the spatial variation is not intensive.  Therefore, we assume that all the reddening values in color bands for the 18 stars are essentially the same. The 18 stars included in each \los\ are sufficient for the Bayesian method according to the Monte Carlo simulations; the average error of reddening is about 0.018 if the number of field stars in each \los\ is less than 18, which is smaller than the photometric error. In total, we get 9,079 valid \los.

To obtain the reddening, we calculate the likelihood values at every point in the parameter grid with a step of 0.005\,mag. In principle, this simple method is too time-consuming to be used on traditional computational devices, especially in the context of high color dimension. Thanks to the up-to-date GPU computational technique that is programmed on a commercial graphics card containing hundreds of cores, this computation can be simultaneously run with hundreds of threads. This powerful technique allows us to calculate the likelihood in the parameter grid directly. In this work, we use NVIDIA Compute Unified Device Architecture (CUDA) to implement the parallel code in the C++ programming language, and running on the GTX580 card, it only takes several seconds to get the posterior distribution of the reddening value spanned in the 4D color space (e.g. $u-g$, $g-r$, $r-i$, and $i-z$) for each \los.

\subsection{Reddening Map}
\begin{figure*}
\centering 
\epsscale{3.0}
\plottwo{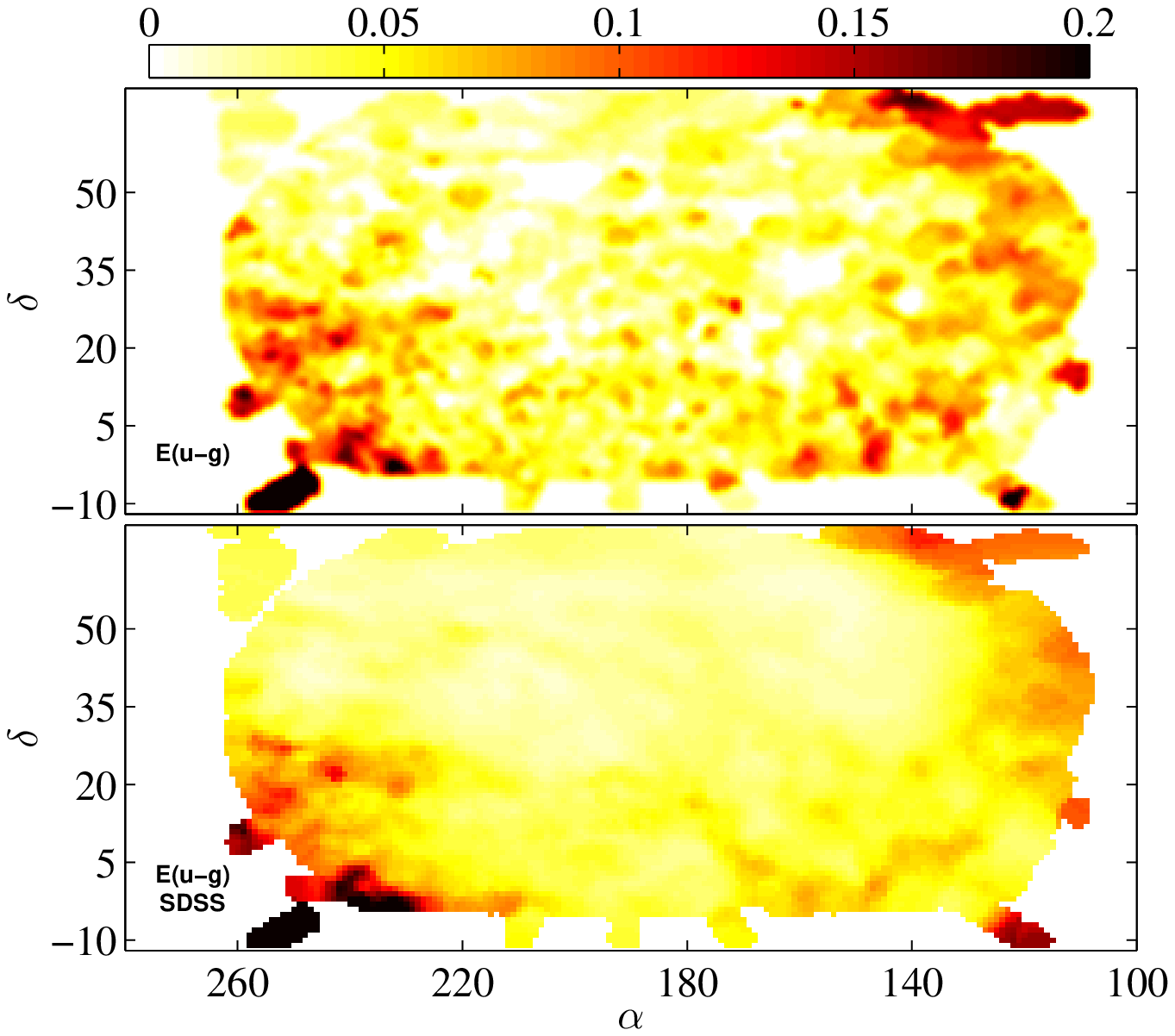}{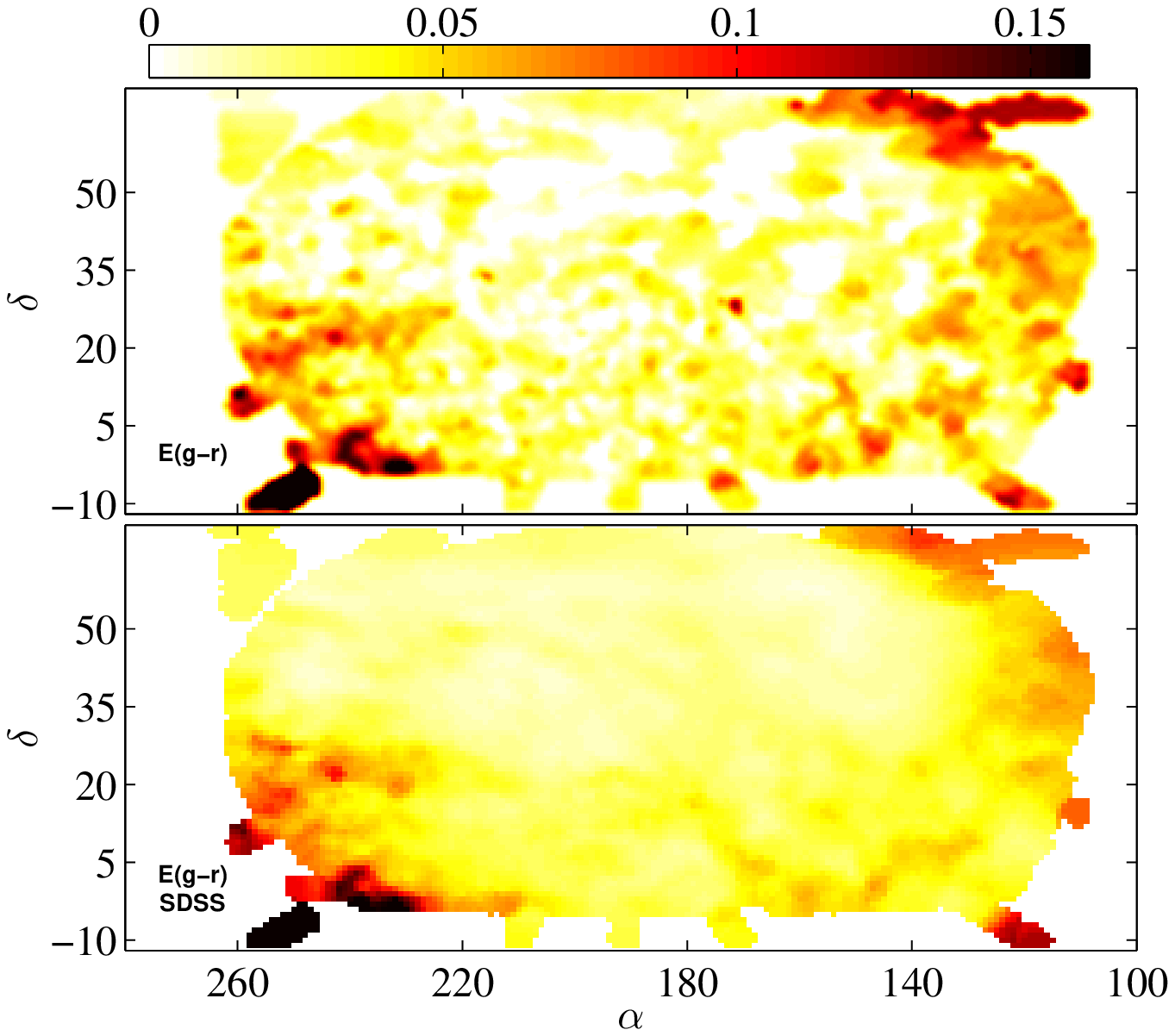}
\caption{The reddening maps for $E(u-g)$ (top panel) and $E(g-r)$ (bottom panel). The top plots in each panel give the reddening map estimated by the Bayesian method, while the bottom ones show the same map released by SDSS. The blanks and other features in our maps have probably occurred because the reddening values are too small at the high Galactic latitudes to be estimated accurately; the reddening values in these regions are very close to the average of photometric error. 
\label{fig_reddening_ugr}}
\end{figure*}

\begin{figure*}
\centering 
\epsscale{3.0}
\plottwo{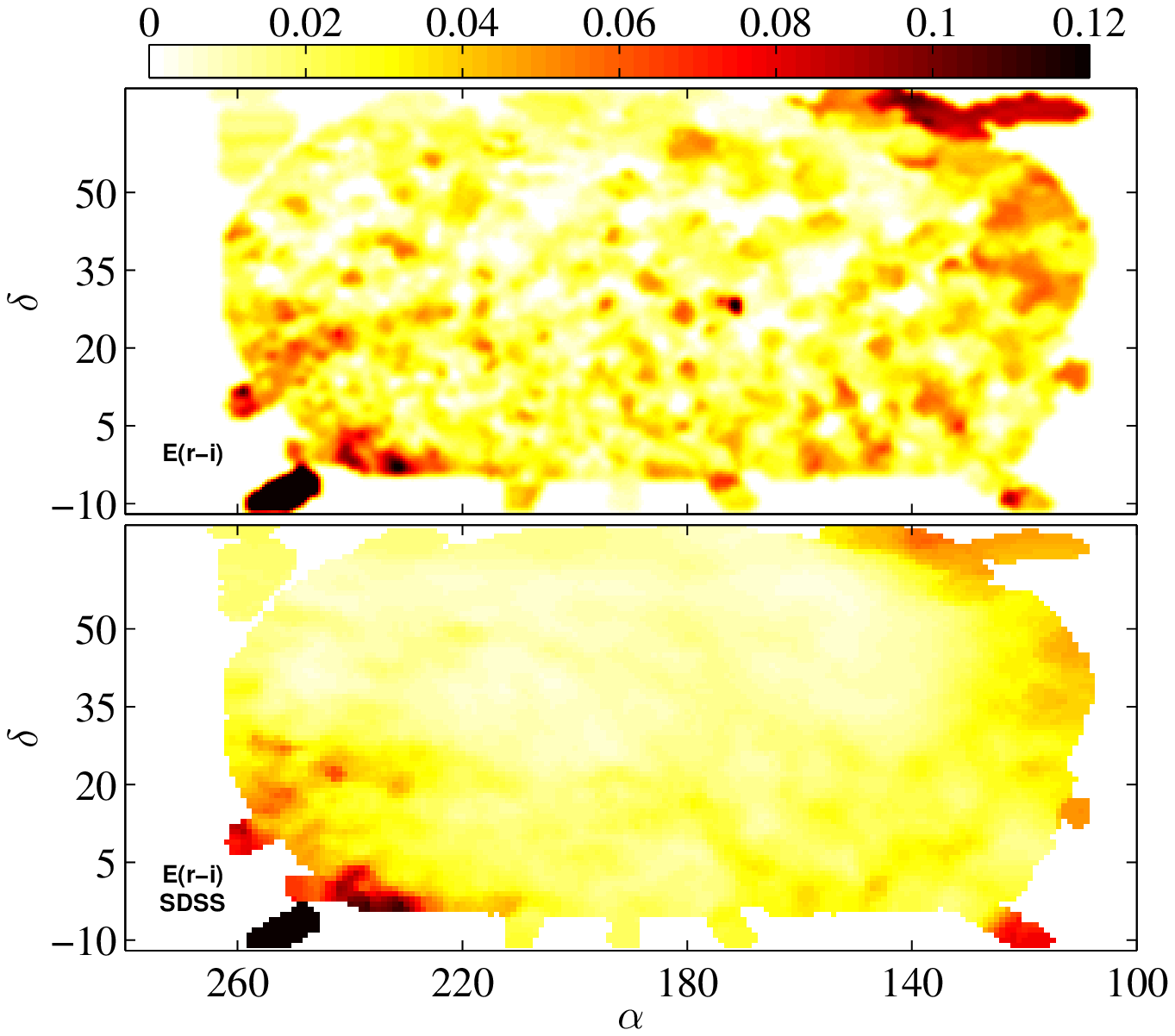}{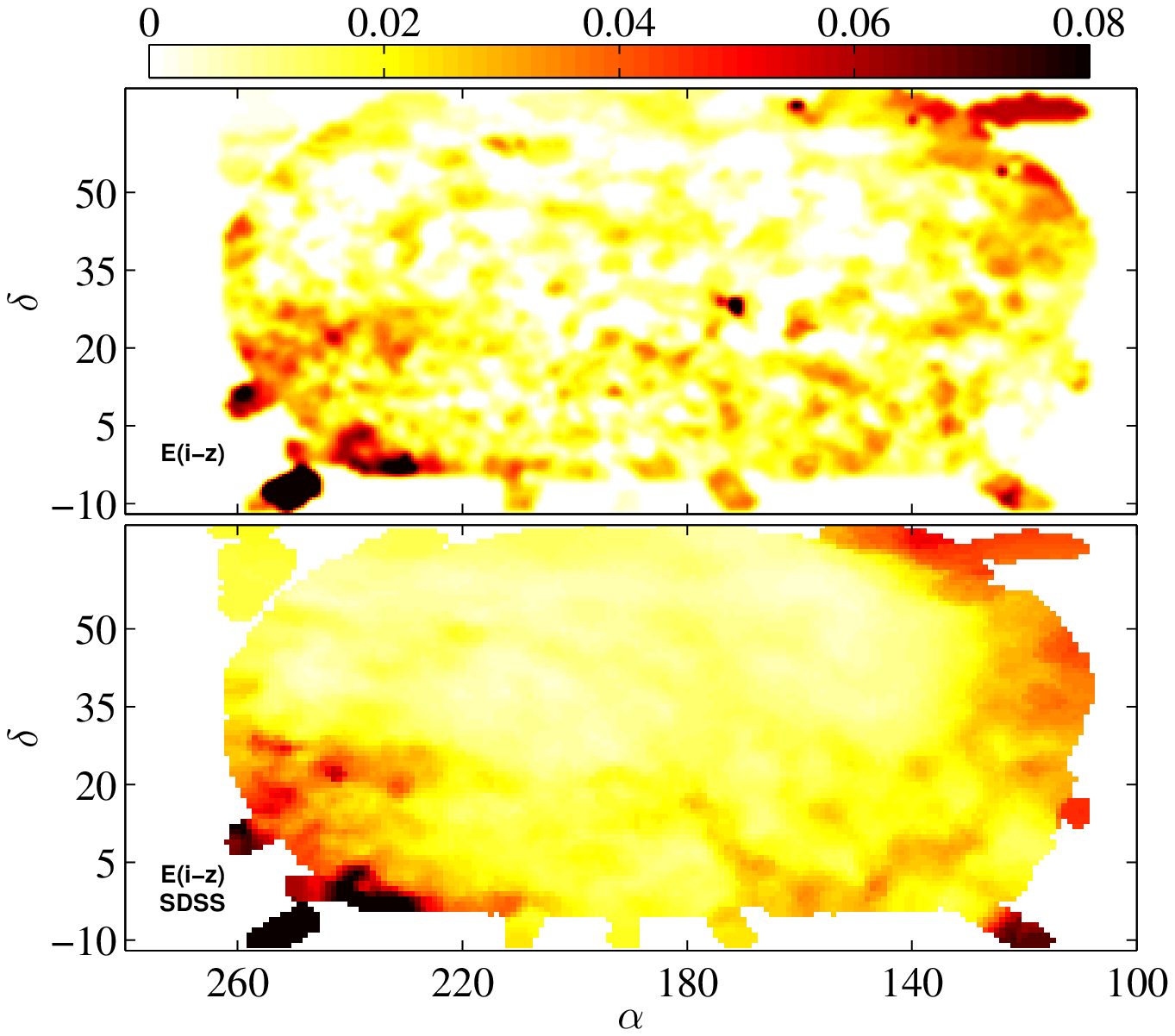}
\caption{The reddening maps for $E(r-i)$ (top panel) and $E(i-z)$ (bottom panel). The top plots in each panel give the reddening map estimated by the Bayesian method, while the bottom ones show the same map released by SDSS.
\label{fig_reddening_riz}}
\end{figure*}

Figure~\ref{fig_reddening_ugr} and~\ref{fig_reddening_riz} present the reddening maps in four colors. The top panels show the results estimated for each color, and the bottom panels are the averaged reddening provided by the SDSS catalogue, that were derived from SFD. 

The blue points scattered in Fig.~\ref{fig_contrast_reddening1} present the differential reddening, $\Delta E(a-b) = E(a-b)_{this} - E(a-b)_{SDSS}$ ($a$ and $b$ stand for any two bands), at variant Galactic latitudes for the four colors. The green lines mark $\Delta E(a-b)=0$, and the red lines are the average of reddening values in Galactic latitude bins. The red lines suggest that the averaged extinctions estimated in this work are very close to those released by SDSS; the $rms$ in the four color are 0.029, 0.022, 0.016, and 0.013, which are almost at the same level as the photometric errors of SDSS.



\begin{figure*}
\centering 
\epsscale{1.5}
\plotone{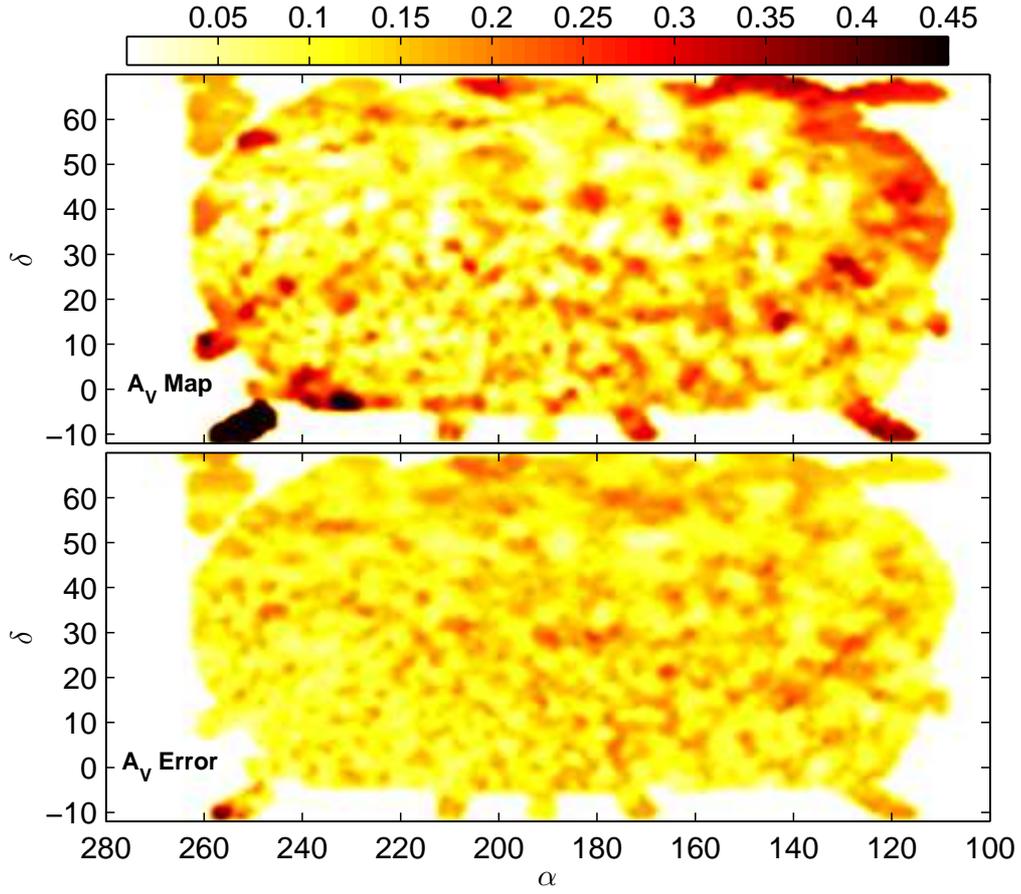}
\caption{The top panel shows the $A_V$ map, while the bottom panel shows its error in the same color level; the value of $A_V$ is the median $A_V$ of all the BHB stars in each \los, and the error is calculated by the median absolute deviation of $A_V$ in each \los. \label{Rv_Av_Map}}
\end{figure*}

To estimate the $R_V$, we adopt the CCM model; $A_V$ and $R_V$ are then estimated according to the method described in Section \ref{method}. Figure~\ref{Rv_Av_Map} shows the filled contour maps of $A_V$ and its error in Galactic coordinates. 

\begin{figure*}
\epsscale{1.5}
\plotone{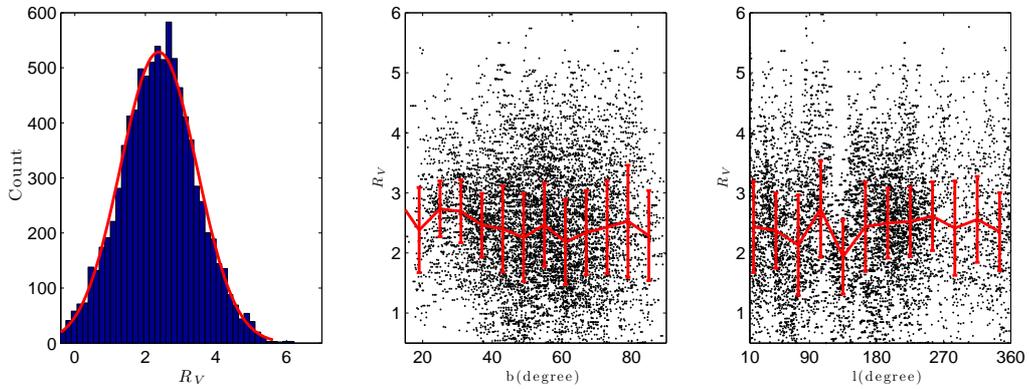}
\caption{The distribution of the estimated $R_V$ is shown in the left panel. The red line shows the best-fit Gaussian with $\mu\simeq 2.4$ and $\sigma\simeq 1.05$. The middle and right panels are the estimated $R_V$ as functions of Galactic latitude and longitude, respectively. The red curves show the $<R_V>$, which keeps constant at $\sim 2.5$ over all latitudes and longitudes.\label{Rv_Hist}
}
\end{figure*}

The left panel in Fig. \ref{Rv_Hist} shows the histogram of $R_V$ (red curve), which is well fitted by a Gaussian with $\mu \sim 2.4$ and $\sigma \sim 1.05$. The middle and right panels are the distribution of $R_V$ at different Galactic latitudes and longitudes, respectively. The red curves show the $<R_V>$, which keeps a constant value of $\sim 2.5$ at all latitudes and longitudes. It implies that $<R_V>$ is invariant with \los\ at high Galactic latitudes. Although the measured $<R_V>$ is smaller than 3.1, which is frequently used, they are in agreement within 1$\sigma$.

\section{Discussion and conclusion}


\cite{smith2010} classifies the BHB and BS stars from their photometry. Thus, the contamination of the BS stars in the BHB catalogue is unavoidable. How do these small fractions of contamination affect the measurement of the extinction?

To investigate the impact of the non-BHB objects, we focus on one \los\ selected randomly, for instance, the \los\ of $(l, b) =(39, 50)$, and calculate the reddening values under the different levels of contamination. We totally get a total of 24 contamination stars in the \los\ of $(l, b) =(39, 50)$ through the following two steps:\\
(1) cross-matching the field BHB stars in the selected \los\ with the catalogue provided by \cite{smith2010}, under the error-radius of $1.0^\circ$; and\\
(2) picking the cross-matched objects with the criteria of $g<19.0$, $P_{SVM}<0.3$ and $modP<0.3$; these two BHB probabilities are the outputs of two different methods of selecting select BHB candidates \citep{smith2010}.

\begin{figure*}
\epsscale{1.6}
\plotone{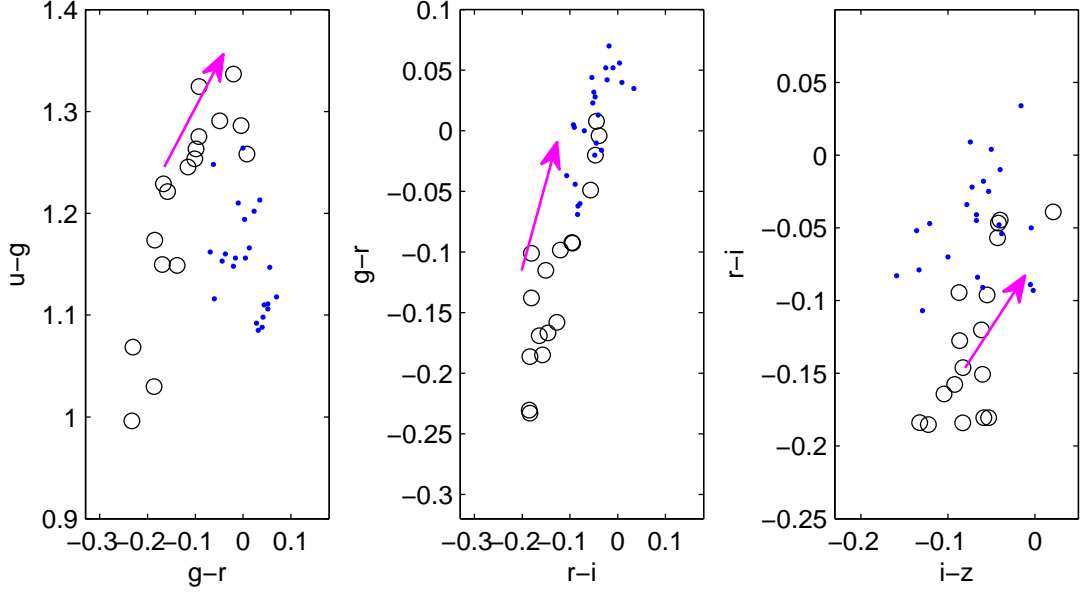}
\caption{Color-color diagrams of the 18 BHB stars (black circles) on the \los\ of $(l, b) =(39, 50)$, and 24 contamination stars (blue points) on this \los, selected from \cite{smith2010}. The magenta arrows present the reddening direction, the starting point of each arrow is the median value of the color indexes of the 18 field stars, and the terminal is the reddened original point with $R_V = 3.1$ and $E(B-V)=0.1$.  
\label{2color_contami}}
\end{figure*}

Figure \ref{2color_contami} shows the color-color diagram of the 24 contamination stars (blue points), with the 18 field BHB stars (black circles); all of these objects are located in the same \los, the \los\ of $(l, b) =(39, 50)$. The red arrows present the reddening direction; the starting point of each arrow is the median value of the color indexes of the 18 field stars; and the terminal is the original point after reddening with the parameters of $R_V = 3.1$ and $E(B-V)=0.1$.

We randomly removed 2, 4, or 6 targets from the 18 BHB field stars, and used the same number of contamination stars, selected arbitrarily from the 24 contaminations, to replace the removed BHB stars. Thus, we could get different levels of contaminations ranging from 11\% to 33\%. Table \ref{tbl-contami} lists the average of reddening values calculated from 50 BHB samples simulated with the different situations of contamination. The results tell us the reddening in the color of $u-g$ is decreased by the increasing contaminations, and the reddening in the color of $g-r$ is slightly increased, while the reddenings are almost not affacted by the contaminations in both the $r-i$ and $i-z$ color indexes. We also ran the simulations of contamination in other \los, for example, the \los\ of $(l,b) = (332, 67)$, and we came to a similar conclusion. The larger $E(u-g)/E(g-r)$ caused by the impact of contamination will give rise to a larger $R_V$ according to Eq. \ref{Extinc}.

In this paper, we use BHB stars to estimate the total extinction in the Milky Way. Because the extinction values are very small at high Galactic latitudes, they require accurate photometry and proper estimation methods. The SDSS photometry provides sufficient accuracy up to 2\%, allowing us to estimate the reddening to an accurate of 0.01 mag. On the other hand, although the BHB stars are sparsely distributed in an elongated region in color space, the Bayesian statistics works well in the determination of the small value of the reddening. Simulations show that by combining the accurate magnitudes of BHB stars with the Bayesian method, we can determine the reddening values in four colors at high Galactic latitudes.

\begin{figure*}
\epsscale{4.0}
\plottwo{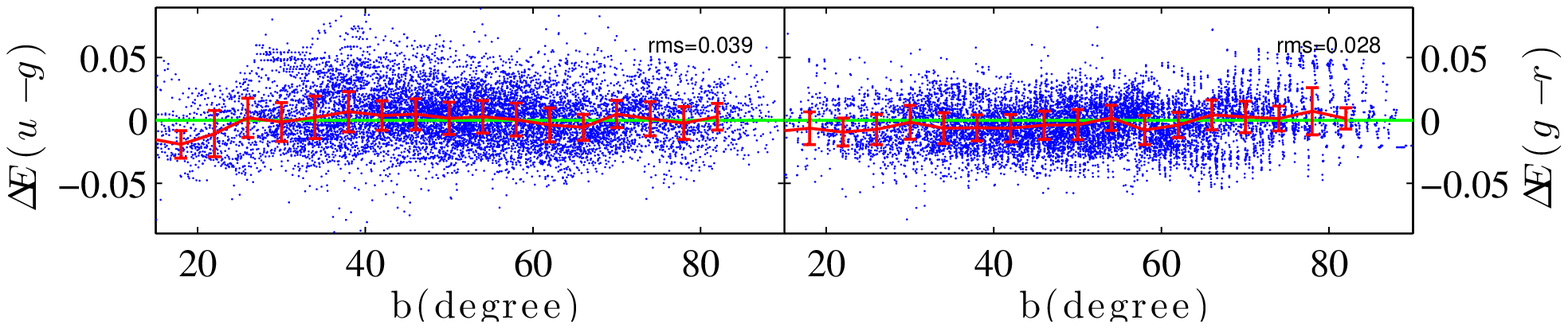}{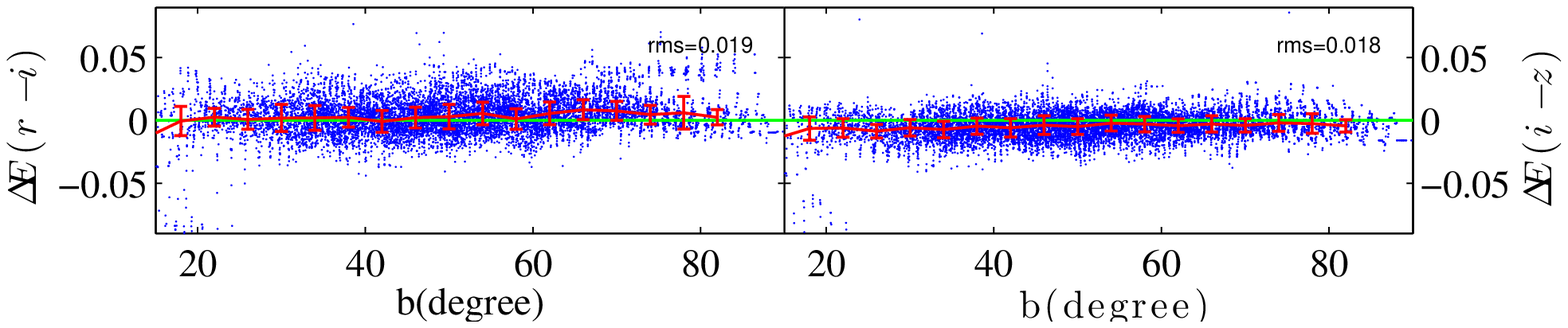}
\caption{The blue dots are the differences between the estimated reddening in this work and those from SDSS at different Galactic latitudes.  The green lines indicate $\Delta E(a-b)=0$ and the red curves the average values of the differential reddening values in different Galactic latitude bins.
\label{fig_contrast_reddening1}}
\end{figure*}

Figure~\ref{fig_contrast_reddening1} shows that the average difference between the reddening measured in this work and SFD is around 0.02\,mag, which approaches to the limit of the uncertainty of the photometry. Therefore, the extinction map derived from the BHB stars are highly consistent with SFD, although there are some slight differences on small scales. 

The $R_V$ inferred from the reddening are centered around 2.4. However, the dispersion of $R_V$ is large. This is partly because the spatial distribution of $R_V$ must not be a single fixed value, but is extended over a range. And the estimation of $R_V$ using very low reddening is crucial and suffers from high uncertainty. Hence, the higher uncertainties eliminate the slight spatial variation of $R_V$ and produce a broad and smooth distribution of the $R_V$.

\begin{acknowledgements}
The authors thank Biwei Jiang and Jian Gao for the helpful discussions, and thank Jaswant Yadav for reading the manuscript. HJT thanks the support from LAMOST Fellowship (No. Y229041001), the China Postdocatoral Science Foundation grant (No. 2012M520384), and the union grant (No. U1231123, U1331202) of NFSC and CAS. CL thanks the support from NSFC grant (No. U1231119), and the 973 Program grant (No. 2014CB845704).  XLC thanks the support from the Ministry of Science and Technology 863 Project grant (No. 2012AA121701), the NSFC grant (No. 11073024, 11103027), and the CAS Knowledge Innovation grant (No. KJCX2-EW-W01). 
\end{acknowledgements}

\onllongtab{1}{
\begin{deluxetable}{ccccccccccccc}
\tabletypesize{\scriptsize}
\tablecaption{Seven globular clusters taken from \citet{harris96}\label{tbl-1}}
\tablewidth{0pt}
\tablehead{
	\colhead{ID}& 
	\colhead{Name} & 
	\colhead{l} & 
	\colhead{b} & 
	\colhead{$R_{sun}$}& 
	\colhead{[Fe/H]}& 
	\colhead{E(B-V)} &
	\colhead{$V_{BH}$} & 
	\colhead{$g_{BH}$\tablenotemark{a}} & 
	\colhead{E(u-g)\tablenotemark{b}}& 
	\colhead{E(g-r)}& 
	\colhead{E(r-i)}& 
	\colhead{E(i-z)} 
}
\startdata
  1 & Pal5 & 0.85 & 45.86 & 23.2 & -1.41 & 0.03 & 17.51 & 17.5 & 0.031  &0.032    &0.019    &0.017\\
  2 & NGC5904 & 3.86 & 46.8 & 7.5 & -1.29 & 0.03 & 15.07 & 15.2 & 0.031    &0.032    &0.019    &0.017\\
  3 & NGC6205 & 59.01 & 40.91 & 7.1 & -1.53 & 0.02 & 14.9 & 15.2 & 0.020&0.021 & 0.013 & 0.011\\
  4 & NGC5466 & 42.15 & 73.59 & 16.0 & -1.98 & 0.016 & 16.52 & 16.6 & 0.016&0.017&0.010& 0.009\\
  5 & NGC5272 & 42.22 & 78.71 & 10.2 & -1.5 & 0.01 & 15.64 & 15.6 & 0.010 &0.011 &0.006 &0.006\\
  6 & NGC5024 & 332.96 & 79.76 & 17.9 & -2.1 & 0.02 & 16.81 & 17.0 &0.020&0.021&0.013&0.011\\
  7 & NGC5053 & 335.7 & 78.95 & 17.4 & -2.27 & 0.01 & 16.69 & 16.8 &0.010&0.011&0.006&0.006\\
\enddata
\tablenotetext{a}{The g-band magnitude of BHB is estimated by averaging over the g magnitude of individual BHB stars located in the corresponding clusters.}
\tablenotetext{b}{The reddening E(a-b) is calculated according to Eq.~\ref{ebv2A} and Eq.~\ref{A2Eab}, given $R_V=3.1$. The differences of E(a-b) is within $<$5\textperthousand \ when the $R_V$ varies from 2.5 and 4.1. Specifically, $\bar{E}(a-b)_{R_V=2.5}=[0.0206,  0.0201,  0.0109,  0.0097]$ and $\bar {E}(a-b)_{R_V=4.1} =[0.0191, 0.0223, 0.0147, 0.0139]$.
}
\end{deluxetable}

\begin{deluxetable}{ccccccccccccccccc}
\tabletypesize{\scriptsize}
\tablecaption{The impact of different ratios of the contamination stars on the reddening\label{tbl-contami}}
\tablewidth{0pt}
\tablehead{
	 \colhead{ID} & 
	\colhead{$E_{ug}$}& 
	 & 
	 &
	& 
	\colhead{$E_{gr}$} & 
	&
	& 
	 & 
	 
	\colhead{$E_{ri}$}&
	& 
	 & 
	& 
       
      \colhead{$E_{iz}$}&
	\\
	& 
	\colhead{$0\%$} & 
	\colhead{$11\%$} & 
	\colhead{$22\%$} & 
	\colhead{$33\%$} &

	\colhead{$0\%$} & 
	\colhead{$11\%$} & 
	\colhead{$22\%$} & 
	\colhead{$33\%$} &

	\colhead{$0\%$} & 
	\colhead{$11\%$} & 
	\colhead{$22\%$} & 
	\colhead{$33\%$} &

	\colhead{$0\%$} & 
	\colhead{$11\%$} & 
	\colhead{$22\%$} & 
	\colhead{$33\%$} 

}
\startdata
 mean & 0.07 & 0.05 & 0.04 & 0.03 & 0.06 & 0.06 & 0.07 & 0.08 & 0.03 & 0.03 & 0.03 & 0.04 & 0.02 & 0.02 & 0.02 & 0.02 \\
\enddata
\tablecomments{This table shows the reddening values on the \los\ of $(l, b) =(39, 50)$, with different ratios of contamination stars. }

\end{deluxetable}

\clearpage

\begin{deluxetable}{ccrrrrrccccc}
\tabletypesize{\scriptsize}
\tablecaption{The 94 template BHB stars\label{tbl-2}}
\tablewidth{0pt}
\tablehead{
  \colhead{ID} &
  \colhead{l} &
  \colhead{b} &
  \colhead{u-g} &
  \colhead{g-r} &
  \colhead{r-i} &
  \colhead{i-z} &
  \colhead{$\sigma(u-g)$} &
  \colhead{$\sigma(g-r)$} &
  \colhead{$\sigma(r-i)$} &
  \colhead{$\sigma(i-z)$} &
  \colhead{GroupID\tablenotemark{a}} 
}
\startdata
  1 & 335.347 & 78.903 & 1.269 & -0.101 & -0.128 & -0.043 & 0.023 & 0.022 & 0.019 & 0.025 & 7\\
  2 & 335.683 & 78.851 & 1.013 & -0.304 & -0.172 & -0.116 & 0.024 & 0.022 & 0.020 & 0.027 & 7\\
  3 & 335.487 & 79.012 & 1.150 & -0.204 & -0.193 & -0.097 & 0.035 & 0.031 & 0.026 & 0.032 & 7\\
  4 & 335.652 & 79.022 & 1.145 & -0.219 & -0.183 & -0.085 & 0.035 & 0.031 & 0.026 & 0.033 & 7\\
  5 & 335.925 & 78.905 & 1.280 & -0.145 & -0.003 & -0.100 & 0.027 & 0.024 & 0.027 & 0.029 & 7\\
  6 & 332.743 & 79.930 & 1.192 & -0.022 & -0.074 & 0.003 & 0.032 & 0.028 & 0.019 & 0.025 & 6\\
  7 & 333.149 & 79.901 & 1.140 & -0.187 & -0.214 & -0.127 & 0.033 & 0.028 & 0.020 & 0.027 & 6\\
  8 & 333.227 & 79.893 & 0.921 & -0.269 & -0.213 & -0.157 & 0.034 & 0.028 & 0.020 & 0.029 & 6\\
  9 & 333.256 & 79.899 & 1.169 & -0.158 & -0.155 & -0.090 & 0.033 & 0.028 & 0.020 & 0.026 & 6\\
  10 & 333.430 & 79.924 & 1.006 & -0.205 & -0.211 & -0.066 & 0.033 & 0.028 & 0.020 & 0.026 & 6\\
  11 & 333.621 & 79.884 & 1.199 & -0.112 & -0.127 & -0.071 & 0.034 & 0.026 & 0.030 & 0.031 & 6\\
  12 & 333.877 & 79.904 & 1.181 & -0.048 & -0.090 & -0.059 & 0.033 & 0.026 & 0.030 & 0.030 & 6\\
  13 & 333.854 & 79.812 & 1.198 & -0.108 & -0.106 & -0.080 & 0.034 & 0.026 & 0.030 & 0.031 & 6\\
  14 & 334.087 & 79.837 & 1.161 & -0.107 & -0.102 & -0.056 & 0.034 & 0.026 & 0.030 & 0.030 & 6\\
  15 & 330.793 & 79.909 & 1.127 & -0.171 & -0.157 & -0.154 & 0.029 & 0.024 & 0.023 & 0.033 & 6\\
  16 & 331.905 & 79.802 & 1.236 & -0.204 & -0.123 & -0.091 & 0.036 & 0.032 & 0.022 & 0.030 & 6\\
  17 & 331.794 & 79.716 & 1.016 & -0.276 & -0.214 & -0.087 & 0.036 & 0.032 & 0.022 & 0.031 & 6\\
  18 & 332.337 & 79.766 & 1.240 & -0.208 & -0.170 & -0.077 & 0.037 & 0.032 & 0.022 & 0.030 & 6\\
  19 & 332.157 & 79.664 & 1.234 & -0.123 & -0.121 & -0.060 & 0.036 & 0.032 & 0.022 & 0.029 & 6\\
  20 & 332.320 & 79.695 & 0.909 & -0.254 & -0.242 & -0.094 & 0.039 & 0.029 & 0.027 & 0.031 & 6\\
  21 & 332.327 & 79.686 & 1.179 & -0.083 & -0.132 & -0.053 & 0.039 & 0.028 & 0.027 & 0.026 & 6\\
  22 & 332.462 & 79.698 & 1.043 & -0.209 & -0.197 & -0.107 & 0.039 & 0.029 & 0.027 & 0.028 & 6\\
  23 & 332.499 & 79.708 & 1.133 & -0.164 & -0.196 & -0.034 & 0.039 & 0.029 & 0.027 & 0.027 & 6\\
  24 & 332.558 & 79.714 & 1.160 & -0.131 & -0.197 & -0.051 & 0.039 & 0.029 & 0.027 & 0.027 & 6\\
  25 & 332.636 & 79.679 & 1.069 & -0.180 & -0.214 & -0.107 & 0.039 & 0.029 & 0.027 & 0.028 & 6\\
  26 & 332.717 & 79.658 & 1.142 & -0.154 & -0.191 & -0.036 & 0.039 & 0.029 & 0.028 & 0.028 & 6\\
  27 & 332.691 & 79.627 & 1.193 & -0.071 & -0.165 & -0.027 & 0.039 & 0.028 & 0.027 & 0.026 & 6\\
  28 & 333.027 & 79.676 & 1.108 & -0.158 & -0.220 & -0.082 & 0.039 & 0.029 & 0.027 & 0.028 & 6\\
  29 & 333.069 & 79.655 & 1.108 & -0.158 & -0.243 & -0.081 & 0.039 & 0.029 & 0.027 & 0.028 & 6\\
  30 & 332.471 & 79.487 & 0.972 & -0.288 & -0.198 & -0.158 & 0.036 & 0.031 & 0.030 & 0.035 & 6\\
  31 & 332.402 & 79.937 & 0.963 & -0.254 & -0.213 & -0.088 & 0.039 & 0.029 & 0.023 & 0.027 & 6\\
  32 & 332.269 & 79.897 & 1.161 & -0.110 & -0.150 & -0.037 & 0.040 & 0.028 & 0.023 & 0.026 & 6\\
  33 & 332.460 & 79.926 & 1.137 & -0.147 & -0.156 & -0.086 & 0.040 & 0.029 & 0.023 & 0.027 & 6\\
  34 & 332.476 & 79.868 & 1.242 & -0.143 & -0.112 & -0.061 & 0.040 & 0.028 & 0.023 & 0.025 & 6\\
  35 & 332.317 & 79.813 & 1.233 & -0.250 & -0.163 & -0.072 & 0.040 & 0.029 & 0.023 & 0.027 & 6\\
  36 & 332.391 & 79.806 & 1.202 & -0.302 & -0.168 & -0.088 & 0.042 & 0.029 & 0.023 & 0.027 & 6\\
  37 & 332.435 & 79.784 & 1.175 & -0.343 & -0.207 & -0.108 & 0.040 & 0.029 & 0.023 & 0.028 & 6\\
  38 & 332.875 & 79.872 & 1.235 & -0.217 & -0.153 & -0.041 & 0.040 & 0.029 & 0.023 & 0.027 & 6\\
  39 & 333.563 & 79.752 & 1.180 & -0.189 & -0.142 & -0.081 & 0.028 & 0.021 & 0.019 & 0.027 & 6\\
  40 & 333.586 & 79.749 & 1.202 & -0.143 & -0.104 & -0.105 & 0.028 & 0.021 & 0.019 & 0.026 & 6\\
  41 & 333.550 & 79.701 & 1.217 & -0.055 & -0.089 & -0.071 & 0.028 & 0.020 & 0.019 & 0.025 & 6\\
  42 & 333.474 & 79.678 & 1.169 & -0.183 & -0.161 & -0.140 & 0.029 & 0.021 & 0.019 & 0.027 & 6\\
  43 & 333.410 & 79.651 & 1.115 & -0.151 & -0.169 & -0.061 & 0.028 & 0.021 & 0.019 & 0.027 & 6\\
  44 & 333.540 & 79.672 & 1.207 & -0.163 & -0.151 & -0.118 & 0.029 & 0.021 & 0.019 & 0.027 & 6\\
  45 & 39.931 & 78.656 & 1.184 & -0.042 & -0.049 & -0.072 & 0.034 & 0.031 & 0.028 & 0.033 & 5\\
  46 & 42.427 & 78.865 & 1.292 & -0.196 & -0.128 & -0.093 & 0.022 & 0.021 & 0.023 & 0.026 & 5\\
  47 & 43.480 & 78.708 & 1.277 & -0.152 & -0.074 & -0.088 & 0.027 & 0.023 & 0.020 & 0.021 & 5\\
  48 & 40.486 & 78.916 & 1.185 & -0.035 & -0.101 & -0.065 & 0.029 & 0.028 & 0.025 & 0.024 & 5\\
  49 & 41.114 & 78.707 & 1.168 & -0.091 & -0.150 & -0.087 & 0.029 & 0.027 & 0.019 & 0.021 & 5\\
  50 & 42.928 & 78.836 & 1.217 & -0.084 & -0.106 & -0.050 & 0.020 & 0.023 & 0.020 & 0.017 & 5\\
  51 & 42.878 & 78.821 & 1.222 & -0.157 & -0.150 & -0.083 & 0.020 & 0.023 & 0.020 & 0.018 & 5\\
  52 & 42.721 & 78.806 & 1.229 & -0.099 & -0.107 & -0.071 & 0.020 & 0.023 & 0.020 & 0.018 & 5\\
  53 & 42.656 & 78.585 & 1.209 & -0.106 & -0.053 & -0.073 & 0.023 & 0.019 & 0.027 & 0.029 & 5\\
  54 & 43.044 & 73.786 & 1.200 & -0.150 & -0.130 & -0.068 & 0.034 & 0.021 & 0.020 & 0.023 & 4\\
  55 & 42.349 & 73.717 & 1.219 & -0.144 & -0.136 & -0.099 & 0.026 & 0.019 & 0.021 & 0.031 & 4\\
  56 & 42.246 & 73.353 & 1.189 & -0.181 & -0.133 & -0.094 & 0.034 & 0.022 & 0.021 & 0.037 & 4\\
  57 & 41.928 & 73.745 & 1.221 & -0.181 & -0.125 & -0.081 & 0.032 & 0.020 & 0.017 & 0.020 & 4\\
  58 & 42.092 & 73.708 & 1.267 & -0.183 & -0.122 & -0.074 & 0.032 & 0.020 & 0.017 & 0.020 & 4\\
  59 & 42.024 & 73.683 & 1.244 & -0.088 & -0.080 & -0.052 & 0.032 & 0.021 & 0.017 & 0.019 & 4\\
  60 & 41.936 & 73.669 & 1.211 & -0.190 & -0.184 & -0.092 & 0.032 & 0.021 & 0.017 & 0.020 & 4\\
  61 & 41.975 & 73.651 & 1.238 & -0.119 & -0.114 & -0.098 & 0.032 & 0.020 & 0.017 & 0.020 & 4\\
  62 & 41.827 & 73.641 & 1.195 & -0.157 & -0.173 & -0.121 & 0.032 & 0.020 & 0.017 & 0.020 & 4\\
  63 & 41.871 & 73.588 & 1.218 & -0.032 & -0.125 & -0.062 & 0.033 & 0.016 & 0.015 & 0.017 & 4\\
  64 & 41.926 & 73.566 & 1.124 & -0.202 & -0.206 & -0.102 & 0.033 & 0.017 & 0.015 & 0.019 & 4\\
  65 & 41.842 & 73.549 & 1.211 & -0.188 & -0.165 & -0.118 & 0.033 & 0.016 & 0.015 & 0.019 & 4\\
  66 & 41.865 & 73.703 & 1.219 & -0.231 & -0.137 & -0.104 & 0.030 & 0.027 & 0.021 & 0.022 & 4\\
  67 & 41.733 & 73.639 & 1.148 & -0.217 & -0.178 & -0.102 & 0.030 & 0.027 & 0.021 & 0.022 & 4\\
  68 & 41.643 & 73.497 & 1.241 & -0.060 & -0.075 & -0.041 & 0.030 & 0.026 & 0.020 & 0.021 & 4\\
  69 & 41.545 & 73.447 & 1.171 & -0.165 & -0.170 & -0.108 & 0.030 & 0.026 & 0.020 & 0.022 & 4\\
  70 & 59.088 & 41.118 & 1.099 & -0.200 & -0.206 & -0.134 & 0.022 & 0.018 & 0.026 & 0.030 & 3\\
  71 & 58.813 & 40.842 & 1.371 & -0.250 & -0.124 & -0.197 & 0.066 & 0.058 & 0.083 & 0.105 & 3\\
  72 & 58.745 & 41.076 & 1.112 & -0.218 & -0.175 & -0.123 & 0.024 & 0.024 & 0.031 & 0.031 & 3\\
  73 & 58.628 & 41.019 & 0.875 & -0.244 & -0.237 & -0.101 & 0.029 & 0.024 & 0.023 & 0.023 & 3\\
  74 & 59.489 & 41.129 & 0.942 & -0.270 & -0.195 & -0.130 & 0.037 & 0.035 & 0.026 & 0.024 & 3\\
  75 & 59.488 & 41.072 & 1.177 & -0.237 & -0.150 & -0.130 & 0.028 & 0.025 & 0.022 & 0.022 & 3\\
  76 & 59.201 & 41.000 & 1.210 & -0.206 & -0.146 & -0.078 & 0.022 & 0.021 & 0.021 & 0.022 & 3\\
  77 & 59.195 & 40.948 & 1.171 & -0.132 & -0.225 & 0.001 & 0.022 & 0.021 & 0.021 & 0.022 & 3\\
  78 & 3.278 & 46.733 & 1.097 & -0.210 & -0.215 & -0.099 & 0.021 & 0.020 & 0.022 & 0.025 & 2\\
  79 & 3.712 & 46.976 & 1.122 & -0.218 & -0.164 & -0.108 & 0.022 & 0.017 & 0.011 & 0.017 & 2\\
  80 & 4.098 & 46.753 & 1.186 & -0.236 & -0.271 & -0.020 & 0.021 & 0.020 & 0.020 & 0.023 & 2\\
  81 & 4.052 & 46.715 & 0.901 & -0.288 & -0.281 & -0.122 & 0.021 & 0.020 & 0.020 & 0.023 & 2\\
  82 & 4.077 & 46.672 & 1.122 & -0.223 & -0.224 & -0.111 & 0.021 & 0.020 & 0.020 & 0.023 & 2\\
  83 & 3.994 & 46.609 & 1.101 & -0.236 & -0.212 & -0.109 & 0.021 & 0.020 & 0.020 & 0.023 & 2\\
  84 & 3.611 & 46.796 & 1.088 & -0.285 & -0.091 & -0.119 & 0.020 & 0.027 & 0.027 & 0.022 & 2\\
  85 & 3.537 & 46.726 & 0.864 & -0.382 & -0.125 & -0.162 & 0.021 & 0.027 & 0.027 & 0.023 & 2\\
  86 & 3.515 & 46.711 & 1.145 & -0.300 & -0.095 & -0.078 & 0.020 & 0.027 & 0.027 & 0.022 & 2\\
  87 & 4.090 & 47.086 & 0.863 & -0.269 & -0.206 & -0.171 & 0.017 & 0.024 & 0.034 & 0.032 & 2\\
  88 & 4.006 & 46.899 & 1.058 & -0.248 & -0.217 & -0.127 & 0.017 & 0.024 & 0.034 & 0.032 & 2\\
  89 & 4.058 & 46.893 & 1.229 & -0.014 & -0.095 & -0.074 & 0.017 & 0.024 & 0.034 & 0.032 & 2\\
  90 & 4.107 & 46.913 & 1.221 & -0.177 & -0.192 & -0.105 & 0.017 & 0.024 & 0.034 & 0.032 & 2\\
  91 & 4.135 & 46.774 & 0.873 & -0.256 & -0.308 & -0.073 & 0.018 & 0.027 & 0.027 & 0.021 & 2\\
  92 & 0.816 & 45.907 & 1.053 & -0.215 & -0.209 & -0.128 & 0.027 & 0.020 & 0.019 & 0.031 & 1\\
  93 & 0.854 & 45.831 & 1.191 & -0.142 & -0.140 & -0.107 & 0.026 & 0.020 & 0.018 & 0.028 & 1\\
  94 & 0.841 & 45.821 & 1.266 & -0.103 & -0.127 & -0.074 & 0.026 & 0.020 & 0.018 & 0.027 & 1\\
\enddata
\tablecomments{The  94 template BHB stars, cross-matched from the seven globular clusters, listed in Table \ref{tbl-2}.}
\tablenotetext{a}{GroupID corresponds to the ID of globular clusters in Table \ref{tbl-2}}
\end{deluxetable}

\clearpage
\begin{deluxetable}{ccrrrrrrrrrrrrcrl}
\tabletypesize{\scriptsize}
\tablecaption{Total extinction in each line of sight (partial)\label{tbl-3}}
\tablewidth{0pt}
\tablehead{
  \colhead{RA} &
  \colhead{DEC} &
  \colhead{$E_{ug}$} &
  \colhead{$E_{gr}$} &
  \colhead{$E_{ri}$} &
  \colhead{$E_{iz}$} &
  \colhead{$\sigma(E_{ug})$} &
  \colhead{$\sigma(E_{gr})$} &
  \colhead{$\sigma(E_{ri})$} &
  \colhead{$\sigma(E_{iz})$} &
  \colhead{$A_V$} &
  \colhead{$\sigma(A_V)$} &
  \colhead{$R_V$} &
  \colhead{$\sigma(R_{V})$} 
}
\startdata
  141.0 & 1.0 & 0.019 & 0.011 & 0.0 & 0.017 & 0.019 & 0.012 & 0.0090 & 0.0080 & 0.105 & 0.171 & 2.374 & 2.716\\
  142.0 & 1.0 & 0.035 & 0.02 & 0.01 & 0.016 & 0.01 & 0.0090 & 0.0090 & 0.0090 & 0.147 & 0.223 & 3.158 & 2.007\\
  143.0 & 1.0 & 0.024 & 0.022 & 0.017 & 0.0080 & 0.0090 & 0.0080 & 0.01 & 0.0090 & 0.152 & 0.14 & 3.626 & 1.869\\
  144.0 & 1.0 & 0.109 & 0.067 & 0.066 & 0.018 & 0.0090 & 0.0080 & 0.0090 & 0.0090 & 0.141 & 0.2 & 2.639 & 1.887\\
  145.0 & 1.0 & 0.122 & 0.088 & 0.068 & 0.032 & 0.0040 & 0.0040 & 0.062 & 0.0050 & 0.209 & 0.167 & 2.639 & 1.575\\
  146.0 & 1.0 & 0.124 & 0.085 & 0.067 & 0.039 & 0.0060 & 0.0080 & 0.0060 & 0.0070 & 0.141 & 0.158 & 2.382 & 1.313\\
  147.0 & 1.0 & 0.125 & 0.092 & 0.049 & 0.033 & 0.0090 & 0.0090 & 0.0090 & 0.0080 & 0.223 & 0.141 & 2.464 & 0.637\\
  148.0 & 1.0 & 0.121 & 0.096 & 0.047 & 0.033 & 0.0070 & 0.0080 & 0.0070 & 0.0080 & 0.192 & 0.095 & 2.554 & 0.654\\
  149.0 & 1.0 & 0.055 & 0.059 & 0.033 & 0.02 & 0.011 & 0.011 & 0.0080 & 0.0090 & 0.15 & 0.074 & 2.554 & 1.065\\
  150.0 & 1.0 & 0.031 & 0.037 & 0.017 & 0.0090 & 0.016 & 0.014 & 0.013 & 0.012 & 0.15 & 0.071 & 2.561 & 1.4\\
  151.0 & 1.0 & 0.038 & 0.034 & 0.021 & 0.014 & 0.011 & 0.01 & 0.0090 & 0.01 & 0.156 & 0.079 & 2.936 & 2.016\\
  152.0 & 1.0 & 0.043 & 0.037 & 0.019 & 0.021 & 0.01 & 0.01 & 0.0090 & 0.0080 & 0.156 & 0.106 & 3.035 & 1.786\\
  153.0 & 1.0 & 0.042 & 0.045 & 0.019 & 0.029 & 0.024 & 0.013 & 0.018 & 0.01 & 0.148 & 0.086 & 3.166 & 2.303\\
  154.0 & 1.0 & 0.053 & 0.044 & 0.026 & 0.03 & 0.018 & 0.012 & 0.015 & 0.01 & 0.148 & 0.086 & 3.166 & 2.21\\
  155.0 & 1.0 & 0.066 & 0.055 & 0.043 & 0.028 & 0.012 & 0.01 & 0.0080 & 0.0090 & 0.204 & 0.124 & 3.673 & 1.739\\
  156.0 & 1.0 & 0.073 & 0.054 & 0.04 & 0.022 & 0.013 & 0.011 & 0.0090 & 0.0080 & 0.226 & 0.14 & 3.493 & 1.669\\
  157.0 & 1.0 & 0.072 & 0.045 & 0.043 & 0.026 & 0.013 & 0.01 & 0.0080 & 0.0080 & 0.204 & 0.142 & 2.991 & 2.241\\
  158.0 & 1.0 & 0.091 & 0.072 & 0.057 & 0.03 & 0.0090 & 0.0060 & 0.0060 & 0.0040 & 0.174 & 0.123 & 3.088 & 1.954\\
  159.0 & 1.0 & 0.109 & 0.064 & 0.053 & 0.012 & 0.011 & 0.0080 & 0.0070 & 0.0090 & 0.124 & 0.144 & 2.216 & 1.89\\
  160.0 & 1.0 & 0.063 & 0.039 & 0.036 & 0.016 & 0.0090 & 0.0090 & 0.0060 & 0.0090 & 0.131 & 0.129 & 2.55 & 1.519\\
  161.0 & 1.0 & 0.038 & 0.041 & 0.028 & 0.011 & 0.01 & 0.012 & 0.0070 & 0.0080 & 0.28 & 0.1 & 2.991 & 1.43\\
  162.0 & 1.0 & 0.045 & 0.029 & 0.02 & 0.021 & 0.01 & 0.016 & 0.01 & 0.0090 & 0.266 & 0.095 & 2.859 & 0.924\\
  163.0 & 1.0 & 0.059 & 0.038 & 0.028 & 0.026 & 0.013 & 0.021 & 0.0090 & 0.0080 & 0.237 & 0.102 & 2.82 & 0.937\\
  164.0 & 1.0 & 0.06 & 0.033 & 0.022 & 0.024 & 0.014 & 0.014 & 0.014 & 0.0090 & 0.131 & 0.124 & 2.555 & 1.78\\
  165.0 & 1.0 & 0.047 & 0.026 & 0.016 & 0.015 & 0.014 & 0.013 & 0.015 & 0.012 & 0.101 & 0.126 & 2.555 & 1.97\\
  166.0 & 1.0 & 0.04 & 0.023 & 0.016 & 0.0050 & 0.012 & 0.013 & 0.014 & 0.011 & 0.114 & 0.106 & 2.555 & 1.97\\
  167.0 & 1.0 & 0.039 & 0.023 & 0.017 & 0.0040 & 0.01 & 0.011 & 0.011 & 0.01 & 0.135 & 0.113 & 2.639 & 1.784\\
  168.0 & 1.0 & 0.046 & 0.031 & 0.026 & 0.0090 & 0.0090 & 0.016 & 0.0090 & 0.01 & 0.142 & 0.106 & 2.201 & 1.548\\
  169.0 & 1.0 & 0.045 & 0.026 & 0.027 & 0.013 & 0.0090 & 0.018 & 0.011 & 0.01 & 0.132 & 0.109 & 2.273 & 1.935\\
  170.0 & 1.0 & 0.057 & 0.046 & 0.034 & 0.02 & 0.014 & 0.014 & 0.0090 & 0.011 & 0.151 & 0.094 & 2.367 & 1.599\\
  171.0 & 1.0 & 0.046 & 0.042 & 0.045 & 0.015 & 0.014 & 0.011 & 0.011 & 0.01 & 0.198 & 0.09 & 3.966 & 1.775\\
  172.0 & 1.0 & 0.046 & 0.04 & 0.049 & 0.016 & 0.011 & 0.01 & 0.0090 & 0.0090 & 0.152 & 0.092 & 3.404 & 1.753\\
  173.0 & 1.0 & 0.048 & 0.035 & 0.042 & 0.022 & 0.018 & 0.015 & 0.014 & 0.01 & 0.159 & 0.125 & 3.181 & 2.32\\
  174.0 & 1.0 & 0.018 & 0.0080 & 0.017 & 0.0080 & 0.01 & 0.017 & 0.0090 & 0.0090 & 0.17 & 0.167 & 3.751 & 2.016\\
  175.0 & 1.0 & 0.022 & 0.015 & 0.019 & 0.0040 & 0.011 & 0.011 & 0.0090 & 0.0090 & 0.271 & 0.194 & 4.452 & 2.336\\
  176.0 & 1.0 & 0.032 & 0.027 & 0.019 & 0.012 & 0.017 & 0.012 & 0.0090 & 0.0090 & 0.166 & 0.188 & 3.724 & 2.83\\
  177.0 & 1.0 & 0.037 & 0.028 & 0.018 & 0.01 & 0.016 & 0.011 & 0.0080 & 0.0090 & 0.166 & 0.195 & 4.126 & 2.75\\
  178.0 & 1.0 & 0.05 & 0.024 & 0.016 & 0.0090 & 0.014 & 0.01 & 0.0080 & 0.0090 & 0.063 & 0.09 & 1.253 & 2.87\\
  179.0 & 1.0 & 0.047 & 0.036 & 0.022 & 0.0070 & 0.012 & 0.011 & 0.0090 & 0.01 & 0.095 & 0.1 & 2.194 & 1.929\\
  180.0 & 1.0 & 0.042 & 0.031 & 0.031 & 0.0090 & 0.012 & 0.011 & 0.0090 & 0.01 & 0.114 & 0.17 & 2.915 & 2.895\\
  181.0 & 1.0 & 0.039 & 0.029 & 0.031 & 0.0090 & 0.01 & 0.01 & 0.0090 & 0.01 & 0.128 & 0.17 & 4.276 & 2.067\\
  182.0 & 1.0 & 0.038 & 0.035 & 0.024 & 0.0060 & 0.0090 & 0.0090 & 0.0080 & 0.0090 & 0.063 & 0.171 & 2.999 & 3.385\\
  183.0 & 1.0 & 0.041 & 0.033 & 0.027 & 0.0050 & 0.0080 & 0.0090 & 0.0070 & 0.0090 & 0.041 & 0.081 & 2.009 & 2.537\\
  184.0 & 1.0 & 0.044 & 0.042 & 0.017 & 0.0070 & 0.0090 & 0.01 & 0.0080 & 0.01 & 0.045 & 0.072 & 1.49 & 1.59\\
  185.0 & 1.0 & 0.041 & 0.035 & 0.026 & 0.0080 & 0.011 & 0.011 & 0.01 & 0.01 & 0.088 & 0.097 & 2.035 & 1.918\\
  186.0 & 1.0 & 0.034 & 0.019 & 0.016 & 0.0080 & 0.014 & 0.014 & 0.0090 & 0.011 & 0.058 & 0.102 & 1.119 & 3.216\\
  187.0 & 1.0 & 0.0060 & -0.021 & 0.013 & -0.016 & 0.01 & 0.01 & 0.0080 & 0.01 & 0.053 & 0.135 & 0.832 & 3.616\\
  188.0 & 1.0 & 0.0060 & -0.015 & 0.0060 & -0.01 & 0.0090 & 0.01 & 0.0080 & 0.0090 & 0.0070 & 0.1 & 0.205 & 3.952\\
  189.0 & 1.0 & 0.042 & 0.017 & 0.02 & 0.0 & 0.012 & 0.011 & 0.01 & 0.012 & 0.035 & 0.106 & 1.138 & 3.282\\
  190.0 & 1.0 & 0.038 & 0.019 & 0.018 & 0.0010 & 0.014 & 0.011 & 0.01 & 0.012 & 0.025 & 0.096 & 0.835 & 1.992\\
  191.0 & 1.0 & 0.043 & 0.026 & 0.024 & 0.0090 & 0.014 & 0.012 & 0.0090 & 0.01 & 0.104 & 0.117 & 1.853 & 1.669\\
  192.0 & 1.0 & 0.054 & 0.018 & 0.024 & 0.0050 & 0.013 & 0.01 & 0.0090 & 0.01 & 0.158 & 0.125 & 2.652 & 1.896\\
  193.0 & 1.0 & 0.059 & 0.015 & 0.023 & 0.0040 & 0.011 & 0.0090 & 0.0080 & 0.0090 & 0.129 & 0.149 & 1.774 & 2.672\\
  194.0 & 1.0 & 0.047 & 0.014 & 0.024 & 0.0 & 0.011 & 0.0090 & 0.0090 & 0.0090 & 0.068 & 0.131 & 1.573 & 2.277\\
  195.0 & 1.0 & 0.034 & -0.014 & 0.013 & -0.011 & 0.011 & 0.01 & 0.0090 & 0.0090 & 0.123 & 0.134 & 1.588 & 1.967\\
  196.0 & 1.0 & 0.026 & -0.0080 & 0.0080 & -0.0070 & 0.01 & 0.012 & 0.011 & 0.0080 & 0.086 & 0.104 & 1.753 & 2.657\\
\enddata
\tablecomments{This table only presents the extinction situation in partial lines of sight; more results can be found on the website.}

\end{deluxetable}

}

\clearpage

\begin{thebibliography}{}

\bibitem[Aihara et al. (2011)]{aihara2011} Aihara, H., Allende, P. C., An, D., Anderson, S. F., et al., 2011, \apjs, 193, 29A
\bibitem[Arce \& Goodman (1999)]{arce99} Arce, H. G. \& Goodman, A. A. 1999, \apj, 512, L135
\bibitem[Berry et al. (2011)]{berry11} Berry, M., Ivezi\'c, Z., Sesar, B., Juri\'c, M., Schlafly, E. F., et al., 2011, \apj, 757, 166B


\bibitem[Cardelli et al. (1989)]{CCM89}  Cardelli, J. A., Clayton, G. C., \& Mathis, J. S. 1989, \apj, 345, 245
\bibitem[Dobashi et al. (2005)]{Dobashi05}Dobashi, K., H. Uehara, R. Kandori, T. Sakurai, M. Kaiden, T. Umemoto, and F. Sato, 2005, PASJ 57, 1.
\bibitem[Draine (2003)]{draine03} Draine, B. T.  2003, \araa,  41, 241

\bibitem[Fang et al. (2011)]{Fang11} Fang, W., Hui, L., M\'enard B., May M., Scranton R., 2011, Phys. Rev. D, 84, 063012
\bibitem[Fermani \& Sch\"onrich (2013)]{Fermani13} Fermani, F., Sch\"onrich, R., 2013, \mnras, 430, 1294F

\bibitem[Fukugita et al. (1996)]{Fukugita96}Fukugita, M., Ichikawa, T., Gunn, J. E., Doi, M., Shimasaku, K.,  Schneider, D. P., 1996, \aj, 111, 1748


\bibitem[Gunn et al. (1998)]{gunn98} Gunn, J. E., Carr, M., Rockosi, C., Sekiguchi, M., et al., 1998, \aj, 116, 3040G
\bibitem[Girardi et al. (2010)]{girardi10}Girardi, L., Williams, B. F., etc. 2010, \apj, 724, 1030
\bibitem[Guy et al. (2010)]{guy10} Guy, J., Sullivan, M., Conley, et al. 2010, A\&A, 537, A7
\bibitem[Harris 1996(2010 edition)]{harris96} Harris, W. E. 1996, \aj, 112, 1487.
\bibitem[Peek \& Graves (2010)]{peek10}Peek, J. E. G., \& Graves, G. J. 2010, \apj, 719, 415
\bibitem[Ross et al. (2011)]{ross11}Ross, A. J., Ho, S, Cuesta, A. J., Tojeiro, R., Percival, W. J., et al., 2011, \mnras, 417, 1350R
\bibitem[Smith et al. (2010)]{smith2010} Smith, K. W., Bailer-Jones, C. A. L., Klement, R. J., Xue, X. X.,  2010, A\&A, 522, A88
\bibitem[Schlegel et al. (1998)]{SFD98} Schlegel, D. J., Finkbeiner, D. P., \& Davis, M. 1998, \apj, 500, 525
\bibitem[Stanek (1998)]{Stanek98} Stanek, K. Z., 1998, ArXiv Astrophysics e-prints arXiv:astro-ph/9802093.
\bibitem[Stoughton et al. (2002)]{Stoughton02}Stoughton, C., et al. 2002, \aj, 123, 485
\bibitem[Schlafly et al. (2010)]{schlafly10} Schlafly, E. F., Finkbeiner, D. P., Schlegel, D. J., et al. 2010, \apj, 725, 1175
\bibitem[Schlafly \& Finkbeiner (2011)]{schlafly11}Schlafly, E. F., Finkbeiner, D. P., 2011, \apj, 737, 103S

\bibitem[Tian et al. (2011)]{tian11}Tian, H. J., Neyrinck, M. C., Budav\'ari, T., Szalay, A. S., 2011, \apj, 728, 34T
\bibitem[Xue et al. (2008)]{xue08}Xue, X. X. et al. 2008, \apj, 684, 1143
\bibitem[Yanny et al. (2000)]{yanny2000} Yanny, B., Newberg, H. J., etc. 2000, \apj, 540, 825.
\bibitem[Yuan et al. (2013)]{yuan13}Yuan, H. B., Liu, X. W., Xiang, M. S., 2013, \mnras, 430, 2188Y

\end{thebibliography}
\end{document}